\def\ln{\ell{n}}
\documentstyle[12pt]{article}
\begin{document}
\begin{titlepage} \vspace{0.2in} \begin{flushright}
MITH-96/16 \\ \end{flushright} \vspace*{0.5cm}
\begin{center} {\LARGE \bf  Quark Masses and Mixing Angles\\} \vspace*{0.8cm}
{\bf She-Sheng Xue}\\ \vspace*{1cm}
INFN - Milan Section, Via Celoria 16, Milan, Italy\\
\vspace*{1.5cm}
{\bf   Abstract  \\ } \end{center} \indent

Inspired by the vector-like phenomenon of chiral gauge theories at short
distances, we postulate that the $W^\pm$-gauge boson has a vector-like gauge
coupling in the high-energy region, beside its purely right-handed gauge
coupling observed in the low-energy region. It is discussed that the top quark
acquires its mass via spontaneous symmetry breaking, while the $u,d,c,s,b$ 
quarks acquire their masses via explicit symmetry breakings. We discover four
relationships between four inter-generation mixing angles and six quark masses
by examining Dyson equations for quark self-energy functions. A preliminary
analysis shows that the hierarchical pattern of quark masses is closely related
to the hierarchical pattern of inter-generation mixing angles. In particular,
the CP-violating phase is predicted to be approximately 79$^\circ$. 

\vfill \begin{flushleft}  September, 1996   \vspace*{2cm} \\
\noindent{\rule[-.3cm]{5cm}{.02cm}} \\
\vspace*{0.2cm} \hspace*{0.5cm} ${}^{a)}$ 
E-mail address: xue@milano.infn.it\end{flushleft} \end{titlepage}

\noindent
{\bf 1.~Motivation}
\vskip0.3cm

The vector-like phenomenon of chiral gauge theories (e.g., the Standard Model)
at short distances, shown by the ``No-Go'' theorem of Nielsen and Ninomiya
\cite{nn81}, is that the fermion spectrum and gauge coupling of regularized
chiral gauge theories must be vector-like if one insists on preserving chiral
gauge symmetries. This seems to run into a paradox that the successful
Standard Model possesses, its very peculiar features of purely
left-handed gauge coupling of $W^\pm$-boson and only left-handed neutrinos.
However, this inconsistency may imply what are the Nature's possible choices 
for the Standard Model at short distances \cite{xue91,ns92,b92}. 

There have been many attempts to find a resolution to this paradox. 
It is currently a very important issue for chiral
gauge theories and the problem has not been completely resolved. We will not
enter into the details of this issue, instead, inspired by this vector-like
phenomenon, we postulate that the effective gauge coupling of $W^\pm$-bosons to
quarks is vector-like in the high-energy region, beside its purely left-handed
gauge coupling observed in the low-energy region: 
\begin{eqnarray}
\Gamma^w_{\mu ij}(q)&=&ig_w(q)\gamma_\mu V_{ij}(P_L+f(q))\label{wv}\\
f(q)&\not=&0,\hskip0.5cm q\sim\Lambda,
\end{eqnarray}
where $g_w(q)$ is a renormalized coupling constant and $V_{ij}$ is the 
Cabibbo-Kobayashi-Maskawa (CKM) \cite{ckm} representation of the
mixing matrix that is parameterized by the following four 
inter-generation mixing angles \cite{pkm},
\begin{equation}
\theta_{12}\hskip0.5cm,
\theta_{23}\hskip0.5cm,
\theta_{13}\hskip0.5cm,
\delta_{13},
\label{4angles}
\end{equation}
where $\theta_{12}=\theta_c$ is the Cabibbo angle and $\delta_{13}$ is the
CP-violating phase. In eq.(\ref{wv}), the non-vanishing of the vertex function
$f(q)$ in the high-energy region $q\sim\Lambda$, which may be the Planck scale,
implies that the gauge coupling (\ref{wv}) is vector-like. Whereas, to
coincide with the parity-violating gauge coupling observed in low-energy
experiments, it
is assumed there is an intermediate energy-threshold $\epsilon$ between the
weak scale ($\sim 250$GeV) and the cutoff $\Lambda$,
\begin{equation}
250GeV <\epsilon < \Lambda,
\label{epsilon}
\end{equation}
at which the
vertex function $f(q)$ vanishes 
\begin{equation}
f(q)|_{q\rightarrow\epsilon}=0.
\label{threshold}
\end{equation}
The gauge coupling (\ref{wv}) seems to violate chiral gauge symmetries.
However, it can be chiral gauge symmetric owing to the possible existence of 
right-handed partners that are three-fermion bound states with appropriate
quantum numbers \cite{xue96} in the high-energy region,
\begin{equation}
(\bar t_R\cdot t_L)t_R,\hskip0.3cm (\bar b_R\cdot b_L)b_R,\hskip0.3cm
\cdot\cdot\cdot,\hskip0.3cm(\bar u_R\cdot u_L)u_R.
\label{threefermions}
\end{equation}
These right-handed
three-fermion bound states disappear at such a threshold $\epsilon$
(\ref{epsilon}), corresponding to the vanishing of $f(\epsilon)$
(\ref{threshold}). We will discuss these three-fermion states 
(\ref{threefermions}), the intermediate energy-threshold (\ref{epsilon}) and 
the vertex function $f(q)$ in 
eq.(\ref{wv}) somewhere else. Fortunately, we do not need the detailed
information of the energy-threshold $\epsilon$ and vertex function $f(p)$ in
the following relevant discussions and preliminary analysis. The
vector-like feature of gauge coupling (\ref{wv}), the intermediate scale
(\ref{epsilon}) and right-handed partners are reminiscent of
``left-right'' symmetric extensions of the Standard model\cite{lr}. 

We stress that our attitude in this paper is that the effective
gauge coupling of the $W^\pm$ bosons (\ref{wv}) is just an {\it assumption}.
The theoretical problems relating to this effective gauge coupling (\ref{wv})
are intentionally not discussed here. Armed with this assumption and
preliminary analysis, we show that the hierarchical pattern of inter-generation
mixing angles 
\begin{equation}
\theta_{12}\gg\theta_{23}\gg\theta_{13},
\label{ha}
\end{equation}
is closely related to the
hierarchical pattern of quark masses
\begin{equation}
m_u\ll\cdot\cdot\cdot\ll m_t.
\label{hb}
\end{equation}
Finally, an approximate prediction of the CP-violating phase ($\delta_{13}\simeq
79^\circ$) is given. 

\vspace*{0.6cm}
\noindent
{\bf 2.~Spontaneous symmetry breaking for the top quark mass}
\vskip0.3cm
\hspace*{0.2cm}
As is known, the gauge interactions in the Standard Model cannot be
the intrinsic dynamics for generating quark masses due to their perturbative
feature. The six Dyson equations for quark self-energy functions $\Sigma_i(p)$
(gap-equations) are self-consistent integral equations that can be, in general,
written in the following form: 
\begin{equation}
\Sigma_i(p)=
\int_{p'}V_i(p,p'){\Sigma_i(p')\over p'^2+\Sigma_i(p')}+m_\circ,
\hskip0.5cm i=u,d,\cdot\cdot\cdot, t\hskip0.3cm {\rm quarks},
\label{gapc}
\end{equation}
where the kernel $V_i(p,p')$ of the integral equations represents the contributions
from all possible gauge interactions in the Standard Model, and $m_\circ$ is an
explicit chiral-symmetry breaking, which is an inhomogeneous term of these
Dyson equations. The notation of internal momentum integration up to the cutoff
$\Lambda$ is defined as, 
\begin{equation}
\int_{p'}=\int_\Lambda {d^4p'\over (2\pi)^4}.
\label{mom}
\end{equation}
In these Dyson-equations, an appropriate gauge (Landau gauge) has been chosen
so that the wave-renormalization $Z_2$ is zero \cite{mn74} in the case of
massless gauge bosons\footnote{As for the case of the massive gauge boson
$Z^\circ$, we consider this to be an approximation owing to the small 
weak-coupling.}. 

When $m_\circ=0$, Dyson equations are homogeneous integral equations, and they
have only trivial solution: 
\begin{equation}
\Sigma_i(p)=0,\hskip0.5cm i=u,d,\cdot\cdot\cdot, t\hskip0.5cm {\rm quarks},
\label{zero}
\end{equation}
for small gauge couplings \cite{mn74,bar1}. It is very important to point out
that $W^\pm$ boson does not contribute to the kernel $V_i(p',p)$ of these Dyson
equations (\ref{gapc}) because of its purely chiral-gauge coupling in the
Standard Model. These six Dyson equations (\ref{gapc}) are decoupled from each
other, namely, there is no mixing between these Dyson equations for
different quarks. 

Bardeen, Hill and Lindner proposed a phenomenal $t\bar t$-condensate model 
\cite{bar}, where a four-fermion interaction of the Nambu-Jona Lasinio type 
\cite{njl} was introduced for the third quark generation $(t, b)$ {\it only},
\begin{equation}
g\bar\psi_L^i(x)\cdot t_R(x)\bar t_R(x)\cdot\psi_L^i(x),
\label{tt}
\end{equation}
with $\psi_L^i(x)$ being the left-handed doublet of the third generation. 
In this case, the kernel in eq.(\ref{gapc}) for the top quark includes the
contribution from the four-fermion coupling beside those from gauge couplings.
When the four-fermion coupling is larger than a certain critical value, 
spontaneous symmetry breaking takes place and the Dyson equation for the top
quark possesses a consistent massive solution even for small gauge
couplings, which can be written as \cite{bar1}. 
\begin{equation}
\Sigma_i(p)\sim m_i\big({p\over m_i}\big)^G,
\hskip0.3cm m_i<p\ll\Lambda,\hskip0.3cm i=t,
\label{solution}
\end{equation}
where $G$ is related to the anomalous dimension of fermion mass operator, which
is a function of perturbative gauge couplings. The top 
quark mass (relating to the weak-scale) is proportional to the condensate 
$\langle\bar t t\rangle$. The fine-tuning of the
four-fermion coupling is necessary so as to have $m_t\ll \Lambda$. 
The $u,d,s,c$ and $b$ quarks remain massless, since the Dyson equations 
(\ref{gapc}) for these quarks are homogeneous and possess only 
trivial solutions (\ref{zero}). 

The Lagrangian being quadratic in fermion fields is one of the prerequisites of the
``No-Go'' theorem for the non-existence of a consistent chiral gauge theory
at short distances. This prerequisite strongly
implies additional four-fermion interactions to the Standard Model at short
distances. One can conceive that these four-fermion interactions are
democratically shared by all quarks, for the reasons that all quarks can be
considered to be massless at the cutoff and the underlying
physics\footnote{That may be the Quantum Gravity.} that induces effective
four-fermion interactions could be flavour-blind. Thus,
unlike the phenomenal $\bar tt$-condensate model (\ref{tt}), we consider that the dynamics of 
four-fermion interactions \cite{njl} should be applied to
all quark flavours with a single four-fermion coupling $g$ at short 
distances,
\begin{equation}
g\sum_{f=1}^3\bar\psi_L^{if}(x)\cdot\psi_R^{jf}(x)\bar\psi_R^{jf}(x)\cdot
\psi_L^{if}(x),
\label{four}
\end{equation}
where the index ``$f$'' is for quark generations and
\begin{eqnarray}
\psi_L^{if}&=&\left(\matrix{u\cr d}\right)_L;
\hskip0.2cm\left(\matrix{c\cr s}\right)_L;
\hskip0.2cm\left(\matrix{t\cr b}\right)_L,
\label{l}\\
\psi_R^{if}&=&\left(\matrix{u_R\cr d_R}\right);
\hskip0.2cm\left(\matrix{c_R\cr s_R}\right);
\hskip0.2cm\left(\matrix{t_R\cr b_R}\right).
\label{r}
\end{eqnarray}
All quark fields are eigenstates of the mass operator. The four-fermion
interactions (\ref{four}) should then be involved in the 
kernel $V_i(p,p')$ of the Dyson equations (\ref{gapc}) for all quarks.

These four-fermion interactions are supposed to undergo the NJL spontaneous
symmetry breaking and the Dyson equations (\ref{gapc}) possess non-trivial solution
(\ref{solution}) for the four-fermion coupling $g$ being larger than its
critical value $(g>g_c)$. One of non-trivial solutions can be
\begin{equation}
\left(\matrix{m_u&0&0\cr 0&m_c&0\cr 0&0&m_t}\right)_{Q={2\over3}};\hskip0.5cm
\left(\matrix{m_d&0&0\cr 0&m_s&0\cr 0&0&m_b}\right)_{Q=-{1\over3}},
\label{max}
\end{equation}
which we call the maximum non-trivial solution\footnote{All condensate 
violating electric charge conservation must be zero.}. In addition, there are 
many possible solutions allowed by the Dyson equations (\ref{gapc}):
\begin{eqnarray}
\left(\matrix{0&0&0\cr 0&0&0\cr 0&0&m_t}\right)_{Q={2\over3}};&\hskip0.5cm&
\left(\matrix{0&0&0\cr 0&0&0\cr 0&0&0}\right)_{Q=-{1\over3}},
\label{min}\\
\left(\matrix{0&0&0\cr 0&m_c&0\cr 0&0&m_t}\right)_{Q={2\over3}};&\hskip0.5cm&
\left(\matrix{0&0&0\cr 0&0&0\cr 0&0&m_b}\right)_{Q=-{1\over3}},
\label{irr}
\end{eqnarray}
{\it etc.}, namely, Dyson equations of some quarks possess trivial solution
(\ref{zero}), which indicates the possibility of some quarks being massive and
others being massless. We call eq.(\ref{min}) the minimum non-trivial solution. 

When these four-fermion interactions undergo the NJL spontaneous symmetry
breaking, quarks acquire mass and the ground states of the whole system
favourably gain negative energy. From this point of view, the maximum
non-trivial solution (\ref{max}) seems to be the ``real'' solution. As a result,
the Dyson equations (\ref{gapc}) engender all quarks to be approximately
equally massive after fine tuning $g\rightarrow g_c^+$, 
\begin{equation}
m_u\simeq m_d\simeq m_s\simeq m_c\simeq m_b\simeq m_t\ll\Lambda,
\label{equal}
\end{equation}
since gauge couplings are perturbatively small, and
corresponding contributions to eqs.(\ref{gapc}) are negligible.
Obviously, this spectrum (\ref{equal}) of equal quark masses is not in
agreement with what is observed. 

However, on the other hand, corresponding numbers of Goldstone and Higgs modes
are produced \cite{gold} when spontaneous symmetry breaking takes place. These
modes carry positive energy in these ground states, and the system gains
positive energy. The more quarks acquire mass, the more such scalar and
pseudoscalar modes are produced. As a consequence, the system should be
stabilized by balancing these two opposite contributions to obtain an
energetically favourable solution. In a lattice model ref.\cite{xue95}, it is
shown that the ground states of the system should be realized by the minimum
non-trivial solution
(\ref{min}) with only one massive quark in the generation that is called $(t,b)$
and three Goldstone modes, 
\begin{eqnarray}
 m_t&\not=&0\nonumber\\
m_u,m_c,m_d,m_s,m_b&= &0,
\label{tb}
\end{eqnarray}
which is actually very close to the real pattern of quark masses.
Those Goldstone modes are to becomes longitudinal modes of the intermediate gauge
bosons $W^\pm$ and $Z^\circ$. There are no extra Goldstone modes. 
This is just the phenomenal $\bar tt$-condensate model proposed in
ref.\cite{bar}. These ground states that are non-perturbatively built by the
phenomenon of spontaneous symmetry breaking should be considered as approximate
ground states, where the approximate pattern (\ref{tb}) is realized. These
arguments based on the energetically favourable realization of the ground
states should be applicable to any models of dynamical symmetry breaking.

\vspace*{0.6cm}
\noindent
{\bf 3. Explicit symmetry breaking for $u,d,s,c,b$ quark masses}
\vskip0.3cm

When $m_\circ\not=0$ in the Dyson equations (\ref{gapc}), there is an explicit
chiral symmetry breaking, these Dyson equations turn out to be
inhomogeneous. For small gauge couplings, 
inhomogeneous Dyson equations for fermion self-energy functions have non-trivial
solutions $\Sigma_i(p)$ stemming from the inhomogeneous term \cite{mn74,bar1}
\begin{equation}
\Sigma_i(p)\sim m_i\big({p\over m_i}\big)^{G'},\hskip0.2cm 
i=u,d,s,c,b,
\label{exp}
\end{equation}
where $G'$ is also related to the anomalous dimension of the fermion mass operator,
and the infrared mass scale $m_i$ is proportional to inhomogeneous terms,
i.e., explicit chiral symmetry breaking $m_\circ$. 

Taking into account the vector-like feature of $W^\pm$ boson coupling
(\ref{wv}) in the high-energy region, we find that this vector-like coupling of
$W^\pm$ boson does contribute to the Dyson equations in the high-energy region:
\begin{eqnarray}
\Sigma_i(p)&=&
\int_{p'} V_{2\over3}(p,p'){\Sigma_i(p')\over p'^2
+\Sigma_i(p')}\nonumber\\
&+&|V_{ij}|^2\int_{|p'|\ge\epsilon} W_{2\over3}(p,p')
{\Sigma_j(p')\over p'^2+\Sigma^2_j(p')},\hskip0.2cm i=u,c,t\hskip0.2cm
{\rm quarks},\label{23}\\
\Sigma_j(p)&=&
\int_{p'} V_{-{1\over3}}(p,p'){\Sigma_j(p')\over p'^2
+\Sigma_j(p')}\nonumber\\
&+&|V_{ji}|^2\int_{|p'|\ge\epsilon} W_{-{1\over3}}(p,p')
{\Sigma_i(p')\over p'^2+\Sigma^2_i(p')}
\hskip0.2cm j=d,s,b\hskip0.2cm {\rm quarks},
\label{13}
\end{eqnarray}
where the six Dyson equations (\ref{gapc}) are classified into two sets of
equations for quarks with charge $Q={2\over3}$ and quarks with charge
$Q=-{1\over3}$ due to the fact that they have different gauge interactions. In
eqs.(\ref{23}) and (\ref{13}), the kernel $W(p,p')$ is proportional to the
vector-like vertex function $f(q)$ and $q=p'-p\sim p'$. The integration of
the internal momentum $p'$ starts from the intermediate threshold $\epsilon$ to the cut-off
$\Lambda$. One can see that eqs.(\ref{23}) and (\ref{13}) are mixed between 
different generations and charge sectors of quarks
via the vector-like vertex function $f(q)$ and the CKM matrix $V_{ij}$. 
Thus, these self-consistent Dyson equations for the six
quark self-energy functions $\Sigma_i(p)$ are coupled together. 

If the top quark mass is generated by spontaneous symmetry breaking, as
discussed in previous section, the Dyson equations for $u,d,s,c,b$ quarks
acquire inhomogeneous terms that are the last terms in eqs.(\ref{23}) and
(\ref{13}).  With respect to the Dyson equations of $u,d,s,c,b$ quarks, these
inhomogeneous terms are explicit chiral symmetry breakings, which are analogous
to $m_\circ$ in eq.(\ref{gapc}). Thus, the Dyson equations of $u,d,s,c,b$
quarks possess non-trivial solutions of the type (\ref{exp}) and $u,d,s,c,b$
quarks are massive. Because these masses are generated by explicit symmetry
breakings, no extra Goldstone bosons are produced. It is worth noting that
possible global symmetries associated with quark generations are explicitly
broken by inhomogeneous terms that contain the CKM matrix, and there are  
no Goldstone bosons associated with these global symmetries. 
 
These inhomogeneous terms, i.e., explicit chiral symmetry breakings are quite
small, since they are proportional to the off-diagonal elements of the CKM
matrix. One can conceive that these small explicit symmetry breakings are
perturbative on the approximate ground states, where the pattern (\ref{tb}) is
realized by the spontaneous symmetry breaking. In other words, when the gauge
couplings and the CKM mixing angles are perturbatively turned on,
spontaneous-symmetry-breaking generated Vacuum alignment must be re-arranged to
the real ground states, where the real pattern is realized. This real pattern
should deviate slightly from the approximate pattern (\ref{tb}), due to the
fact that gauge couplings are perturbatively small and the observed CKM mixing
angles are small deviations from triviality. 

This new alignment of vacua (the ground states) certainly minimizes the energy
in such a way that all quark self-functions $\Sigma_i(p)$ satisfy the six
self-consistently coupled Dyson equations (\ref{23}) and (\ref{13}). These
equations (\ref{23}) and (\ref{13}) show that six quark masses and four CKM
mixing angles are closely related. This means the six quark masses and four CKM
mixing angles are no longer completely free and independent parameters.
Nevertheless, we are still far from entirely determining the quark masses and
CKM mixing angles, since the Dyson equations (\ref{23}) and (\ref{13}) are not
complete equations for the dynamics of the whole system. The best we can
achieve is to find the explicit relationships between the six quark masses and
four CKM mixing angles.

\vspace*{0.6cm}
\noindent
{\bf 4. Preliminary analysis of the Dyson equations}
\vskip0.3cm

The Dyson equations (\ref{23}) and (\ref{13}) in general are very complicated. It
is hard to see any explicit relationships between the six quark masses and four CKM
mixing angles. In order to find the relationships between quark masses and
mixing angles, we have to figure out the most important mass-dependence in these
Dyson equations and make a reasonable approximation. Based on the arguments of
all quarks being massive, which we discussed in previous sections, we adopt the
non-trivial solution (\ref{solution},\ref{exp}) as an ansatz for each quark
self-energy function. Substituting these ansatze into eqs.(\ref{23}) and
(\ref{13}), we obtain approximately 
\begin{eqnarray}
Q^i_{2\over3}(p,\{m\})m^i_{2\over3}=\Sigma_jK^j_{-1\over 3}(p,\{m\})
|V_{ij}|^2m^j_{-{1\over3}},\hskip1cm
i=u,c,t\label{1}\\
Q^j_{-{1\over3}}(p,\{m\})m^j_{-{1\over3}}=\Sigma_i
K^i_{2\over 3}(p,\{m\})|V_{ji}|^2m^i_{2\over3},\hskip1cm
j=d,s,b,
\label{2}
\end{eqnarray}
where the equations are factorized with 
\begin{eqnarray}
m^i_{2\over3}&=&m_u,m_c,m_t\nonumber\\
m^j_{-{1\over3}}&=&m_d,m_s,m_b,
\label{quarkmass}
\end{eqnarray}
being the quark
masses for the $Q={2\over3}$ and $Q=-{1\over3}$ sectors respectively.
The factors $Q^i_{2\over3}(p,\{m\})$, $Q^j_{-{1\over3}}(p,\{m\})$,
$K^i_{2\over3}(p,\{m\})$ and $K^j_{-{1\over3}}(p,\{m\})$ in
(\ref{1},\ref{2}) are still quite complicated, 
\begin{eqnarray}
Q^i_{2\over3}(p,\{m\})&=&({p\over m_i})^G-
\int_{p'}V^i_{2\over3}(p,p'){({p'\over m_i})^G\over p'^2+m_i
({p'\over m_i})^G},\hskip0.2cm i=u,c,t,\label{q23}\\
Q^j_{-{1\over3}}(p,\{m\})&=&({p\over m_j})^G-
\int_{p'} V^j_{-{1\over3}}(p,p'){({p'\over m_j})^G\over p'^2+m_j
({p'\over m_j})^G},\hskip0.2cm j=d,s,b\label{q13}
\end{eqnarray}
and 
\begin{eqnarray}
K^i_{2\over 3}(p,\{m\})&=&\int_{|p'|\ge\epsilon} 
W_{2\over3}(p,p'){({p'\over m_i})^G\over p'^2+m_i
({p'\over m_i})^G},\hskip0.3cm i=u,c,t,
\label{k23}\\
K^j_{-1\over 3}(p,\{m\})&=&\int_{|p'|\ge\epsilon} W_{-1\over3}(p,p')
{({p'\over m_j})^G\over p'^2+m_j
({p'\over m_j})^G},\hskip0.3cm j=d,s,b,
\label{k13}
\end{eqnarray}
where 
\begin{equation}
\{m\}=m_u,m_d,m_s,m_c,m_b,m_t.
\label{m}
\end{equation}
Obviously, these functions $Q$ and $K$ depend on all quark masses (\ref{m})
through renormalizations. However, these mass-dependences are logarithmically
weak $(\ln{\Lambda\over m_i})$ for renormalizable gauge interactions. These can
be seen by dimensional counting of integrations in eqs.(\ref{q23})-(\ref{k13}),
in which functions $V^i(p',p)$ and $W(p',p)$ contain propagators of gauge
bosons and behave as $O({1\over p'^2})$ in the high-energy region
($p'\rightarrow\Lambda$). The contributions of NJL four-fermion interactions to
functions $Q(p,\{m\})$ are 
\begin{equation}
1-{8\pi^2\over N_cg_c\Lambda^2}-\left({m_i\over\Lambda}\right)^2\ln{\Lambda\over m_i}
+O\left[\left(\ln{\Lambda\over m_i}\right)^2\right], 
\label{njlgap}
\end{equation}
which is also logarithmically dependent on quark masses $\{m\}$ after 
fine-tuning $g\rightarrow g_c^+$ to cancel quadratic divergent terms.

On the basis of the factors $Q^i_{2\over3}(p,\{m\})$,
$Q^j_{-{1\over3}}(p,\{m\})$, $K^i_{2\over 3}(p,\{m\})$ and $K^j_{-1\over
3}(p,\{m\})$ logarithmically depending on quark masses, which are
very weakly dependent in comparison with the linear mass-dependence 
in eqs.(\ref{1}) and (\ref{2}), we can make the completely reasonable approximations: 
\begin{eqnarray}
Q^u_{2\over3}&\simeq& Q^c_{2\over3}\simeq Q^t_{2\over3};\hskip1cm    
Q^d_{-{1\over3}}\simeq Q^s_{-{1\over3}}\simeq Q^b_{-{1\over3}},\nonumber\\
K^u_{2\over3}&\simeq& K^c_{2\over3}\simeq K^t_{2\over3};\hskip1cm    
K^d_{-{1\over3}}\simeq K^s_{-{1\over3}}\simeq K^b_{-{1\over3}},
\label{qq}
\end{eqnarray}
in eqs.(\ref{1}) and (\ref{2}).
This means that the variations of the functions $Q$ and $K$ in terms of
different quark masses are negligible in these equations. It is important to
note that the factors $Q^i_{2\over3}(p,\{m\})$ and $Q^j_{-{1\over3}}(p,\{m\})$
are different due to different gauge interactions. 

Taking the ratios from the six Dyson equations (\ref{1},\ref{2}), the 
functions $Q$ and $K$ are approximately cancelled and we
obtain the following equations: 
\begin{eqnarray}
{m_u\over m_c}&=&{|V_{ud}|^2m_d+|V_{us}|^2m_s+|V_{ub}|^2m_b\over
|V_{cd}|^2m_d+|V_{cs}|^2m_s+|V_{cb}|^2m_b},\label{r1}\\
{m_u\over m_t}&=&{|V_{ud}|^2m_d+|V_{us}|^2m_s+|V_{ub}|^2m_b\over
|V_{td}|^2m_d+|V_{ts}|^2m_s+|V_{tb}|^2m_b},\label{r2}\\
{m_c\over m_t}&=&{|V_{cd}|^2m_d+|V_{cs}|^2m_s+|V_{cb}|^2m_b\over
|V_{td}|^2m_d+|V_{ts}|^2m_s+|V_{tb}|^2m_b},\label{r1'}
\end{eqnarray}
and
\begin{eqnarray}
{m_d\over m_s}&=&{|V_{du}|^2m_u+|V_{dc}|^2m_c+|V_{dt}|^2m_t\over
|V_{su}|^2m_u+|V_{sc}|^2m_c+|V_{st}|^2m_t},\label{r3}\\
{m_d\over m_b}&=&{|V_{du}|^2m_u+|V_{dc}|^2m_c+|V_{dt}|^2m_t\over
|V_{bu}|^2m_u+|V_{bc}|^2m_c+|V_{bt}|^2m_t},\label{r4}\\
{m_s\over m_b}&=&{|V_{su}|^2m_u+|V_{sc}|^2m_c+|V_{st}|^2m_t\over
|V_{bu}|^2m_u+|V_{bc}|^2m_c+|V_{bt}|^2m_t}.\label{r3'}
\end{eqnarray}
There are only four independent equations that completely determine the 
four CKM mixing angles in terms of six quark masses. 

Using the standard parameterization of the CKM matrix \cite{pkm} with four
mixing angles (\ref{4angles}), one can obtain definite, though rather complicated
relationships between such mixing angles and the pattern of quark masses. 
Defining, for convenience
\begin{eqnarray}
x&=&\tan^2\theta_c;\hskip1cm y=\sin^2\theta_{23}\nonumber\\
z&=&\cot^2\theta_{13};\hskip1cm w=\cos\delta_{13},\nonumber
\end{eqnarray}
we cast eqs.(\ref{r1},\ref{r2},\ref{r3},\ref{r4}) into,
\begin{eqnarray}
{m_u\over m_c}&=&{z(m_d+xm_s)+(1+x)m_b\over (xm_d{+}m_s)(1{-}y)(1{+}z)
{+}y(m_d{+}xm_s){+}(1{+}x)yzm_b{+}(m_d{-}m_s)wc},\label{cos}\\
{m_u\over m_t}&=&{z(m_d+xm_s)+(1+x)m_b\over y(1{+}z)(xm_d{+}m_s){+}(1{+}y)
[m_d{+}xm_s
{+}(1{+}x)zm_b]{+}(m_s{-}m_d)wc},\label{tan}\\
{m_d\over m_s}&=&{zm_u{+}[(1{+}z)x(1{-}y){+}y]m_c{+}[xy(1{+}z){+}1{-}y]m_t{+}
(m_c{-}m_t)wc\over
xzm_u{+}[(1{-}y)(1{+}z){+}xy]m_c{+}[y(1{+}z){+}x(1{-}y)]m_t{+}(m_t{-}m_c)wc},\label{sin}\\
{m_d\over m_b}&=&{zm_u{+}[(1{+}z)x(1{-}y){+}y]m_c{+}[xy(1{+}z){+}1{-}y]m_t{+}
(m_c{-}m_t)wc\over
(1+y)[m_u+z(ym_c+(1-y)m_t)]},
\label{cot}
\end{eqnarray}
where
\begin{equation}
c=2\sqrt{xy(1-y)(1+z)}.
\label{c}
\end{equation}
These equations still look very complicated.

The relationships (\ref{r1}), (\ref{r2}), (\ref{r3}) and (\ref{r4}) still cannot 
tell us the observed hierarchical pattern of quark masses. However,
we should emphasize at this point that in the Standard Model, the CKM mixing
angles are totally extrinsic elements, qualifying the observed pattern as real
as compared (and opposed) to any other {\it possible} pattern, where the CKM
mixing angles can be either trivial (the {\it approximate} pattern (\ref{tb}))
or anything one wishes. We consider the observed {\it real} pattern so close to
the {\it approximate} one (\ref{tb}), this implies in turn that we do live in a
world where the CKM matrix is {\it almost} trivial. 

The equations (\ref{cos})-(\ref{cot}) can be drastically simplified if one takes into
account the observed hierarchical pattern of quark masses, i.e., the fact that 
the top quark is much heavier than the others. One can then approximately 
solve the above four independent equations (\ref{cos})-(\ref{cot}) and get
\begin{eqnarray}
\tan^2\theta_c&\simeq& -{m_d\over m_s};\hskip1cm\sin^2\theta_{23}\simeq
-{m_c\over 2m_t},\label{res1}\\
\tan^2\theta_{13}&\simeq& {m_u\over m_t};\hskip1cm \cos\delta_{13}\simeq
\sqrt{m_sm_u\over 2m_cm_d},\label{res2}
\end{eqnarray}
where the strange and ominous looking minus signs, can all be eliminated by
chirally rotating the
fields of the (c, s) generation so as to transform 
\begin{equation}
-m_s, -m_c\rightarrow m_s,m_c.
\label{sign}
\end{equation}
We note that
\begin{equation}
\tan\theta_c\simeq{m_d\over m_s}
\label{cabb}
\end{equation}
in eq.(\ref{res1}) has been around in completely different contexts for a
quarter of a century\cite{f} now, while the others\footnote{I thank 
Dr.~Z.~Z.~Xing for his comments and bringing my attention to the 
ref.\cite{fx}} -- to our knowledge -- appear
to be genuine consequences of the peculiar chiral symmetry breaking of both the
spontaneousness and explicitness, which we have fully discussed in 
sections 3 and 4.
The most interesting aspect of eqs.(\ref{res1}) and (\ref{res2}) is their tying
the observed hierarchical pattern of the inter-generation mixing angles 
(\ref{ha}) to the observed hierarchy of mass ratios
\begin{equation}
{m_s\over m_d}\gg{m_c\over 2m_t}\gg{m_u\over m_t}.
\label{hm}
\end{equation}
Finally, as for the
CP-violating phase $\delta_{13}$, eqs.(\ref{r1}), (\ref{r2}), (\ref{r3}) and 
(\ref{r4}) can be cast in the form 
\begin{equation}
\cos\delta_{13}\simeq{1\over2}{\sin\theta_{13}\cos\theta_{23}\cos\theta_c\over
\sin\theta_{23}\sin\theta_c}, 
\label{cp} 
\end{equation} 
which by using the
experimental determinations of the mixing angles (central values
$\sin\theta_c=0.221; \sin\theta_{23}=0.040; \sin\theta_{13}=0.0035$\cite{data})
yields the prediction 
\begin{equation}
\delta_{13}\simeq 79^\circ.
\label{cpa}
\end{equation}

In summary, starting with the postulation of the vector-like feature of the
$W^\pm$ boson gauge coupling in the high-energy region, we study the Dyson
equations for the six quark self-energy functions. We discuss the possible pattern
of quark masses on the basis that the top quark mass is generated by
NJL spontaneous symmetry breaking and $(u,d,s,c,b)$ masses are generated by
explicit chiral symmetry breaking. The four relationships between the six quark
masses and four CKM mixing angles are approximately derived by the preliminary
analysis of the Dyson equations. These four relationships, which relate the
hierarchical pattern of the four CKM mixing angles to the hierarchical pattern 
of the six quark masses, are in good agreement with observations. Of particular interest is
the prediction (\ref{cpa}) of the CP-violating phase. We have to confess that
the practical analysis presented in this article is very preliminary and more
precise analysis needs pursuing further.

\end{document}